\documentclass[runningheads]{llncs}
\usepackage[T1]{fontenc}
\usepackage{graphicx}
\usepackage{booktabs}
\usepackage{rotating}
\usepackage{subcaption}
\usepackage{comment}
\usepackage{hyperref}

\begin{document}
%
\title{PTFA: An LLM-based Agent that Facilitates Online Consensus Building through Parallel Thinking}
\titlerunning{PTFA: Parallel Thinking-based Facilitation Agent}
%
\author{Wen Gu\inst{1^*} \and
Zhaoxing Li\inst{2} \and
Jan Buermann\inst{2} \and Jim Dilkes\inst{2} \and Dimitris Michailidis\inst{3} \and Shinobu Hasegawa \inst{4} \and Vahid Yazdanpanah\inst{2} \and Sebastian Stein\inst{2}}
\authorrunning{W. Gu et al.}
%
\institute{Nagoya Institute of Technology, Nagoya, Japan \\
\email{wgu@nitech.ac.jp} \and
University of Southampton, Southampton, UK \\
\email{\{Zhaoxing.Li,  J.Buermann, J.Dilkes, V.Yazdanpanah\}@soton.ac.uk, ss2@ecs.soton.ac.uk}\\
\and
University of Amsterdam, Amsterdam, Netherlands \\
\email{d.michailidis@uva.nl}
\and
Japan Advanced Institute of Science and Technology, Nomi, Japan\\
\email{hasegawa@jaist.ac.jp}}
\maketitle              
\begin{abstract}
Consensus building is inherently challenging due to the diverse opinions held by stakeholders. 
Effective facilitation is crucial to support the consensus building process and enable efficient group decision making. 
However, the effectiveness of facilitation is often constrained by human factors such as limited experience and scalability. 
In this research, we propose a Parallel Thinking-based Facilitation Agent (PTFA) that facilitates online, text-based consensus building processes. 
The PTFA automatically collects real-time textual input and leverages large language models (LLMs) to perform all six distinct roles of the well-established Six Thinking Hats technique in parallel thinking.
To illustrate the potential of the agent, a pilot study was conducted, demonstrating its capabilities in idea generation, emotional probing, and deeper analysis of idea quality. 
Additionally, future open research challenges such as optimizing scheduling and managing behaviors in divergent phase are identified.
Furthermore, a comprehensive dataset that contains not only the conversational content among the participants but also between the participants and the agent is constructed for future study.

\end{abstract}
\section{Introduction}
Collective intelligence, the emergent capability of groups to solve problems and make decisions collaboratively, has become a cornerstone of effective decision making in diverse domains, from business strategy to public policy \cite{joshua_network_2017}. 
By integrating the knowledge, creativity, and perspectives of multiple individuals, collective intelligence can produce outcomes that surpass those achievable by individuals working in isolation. 
However, harnessing this potential often requires skilled facilitation aimed at supporting decision making, such as guiding the process, promoting constructive interaction and inclusivity, and managing content in a structured manner \cite{gimpel_digital_2024}.
To address the need for effective group consensus building facilitation, artificial intelligence (AI)-based approaches leveraging machine learning and intelligent agents have been proposed to support tasks such as content generation, balancing participation, and accumulating facilitation knowledge~\cite{ito_an_2022,hadfi_conversational_2023,gu_case-based_2021}.
In particular, the advent of generative AI and large language models (LLMs) has demonstrated significant potential in enhancing specific facilitation tasks, including collaborative brainstorming \cite{moeka_towards_2024}, generating and refining statements \cite{Tessler_common_ground} and mitigating conflicts \cite{ma2024towards}. 
Despite recent advances, achieving automated systematic facilitation support remains challenging, particularly because it involves managing multiple facilitation tasks concurrently and managing the relationships between those tasks in dynamic decision-making processes. 
For example, systematic facilitation requires the ability to understand the context of the decision-making process, determine how to promote participation, and adapt accordingly based on the identified characteristics and needs of the participants. 
Additionally, it must determine the appropriate timing for executing these facilitation tasks.
This requires abilities such as sophisticated contextual understanding, maintaining the natural flow of discussions, and striking a delicate balance between structured guidance and flexibility to avoid rigidity.

To bridge the gap between specific facilitation task support and systematic automated facilitation, we propose a novel Parallel Thinking-based Facilitation Agent (PTFA) designed to support online group consensus building.
This conversational agent actively participates in the group decision-making process, guiding stakeholders through structured, context-aware interactions to foster mutual understanding and consensus.
To address multiple facilitation tasks concurrently, PTFA integrates the principles of parallel thinking, a structured methodology for systematically exploring multiple perspectives, with the advanced capabilities of LLMs \cite{gocmen_effects_2019,wan_second_mind}. 
Specifically, PTFA is based on the Six Thinking Hats technique, which is a successful method for finding consensus \cite{de2017six}.
Each hat represents a role of an effective facilitator, e.g., the blue hat represents the role of managing the discussion and the green hat represents fostering creativity.
The roles and suitable discussion interactions are selected automatically and dynamically based on the contributions of the participants.
To evaluate PTFA, we developed an online text-based discussion platform, serving as an interactive interface for participants to gather and conduct discussions. 
We conducted a pilot study involving 16 discussion groups and 48 participants in an online environment collecting a rich dataset on participant interaction with the PTFA which has allowed us to establish guidelines on pursuing AI-based consensus-building platform development in idea generation, emotional probing and deeper argument analysis as well as challenges such as automated phase management and dynamic response timing control.
To summarize, the main contributions of this paper are as follows:
\begin{itemize}
    \item We propose a novel agent-based framework that implements a multi-role facilitation model powered by an LLM, enabling automated and systematic support for group decision-making.
    \item We implement the proposed approach in an online, free-text decision-making scenario and conduct real-world pilot experiments to evaluate its potential in idea generation, emotional probing, and deeper analysis of idea quality and find limitations in optimizing scheduling and managing behaviors as a fully automated facilitator.
    \item We construct a novel dataset rich in conversational interactions, capturing dialogues both among participants and between participants and the agent.
\end{itemize}

The remainder of this paper is organized as follows. 
Section \ref{sec:relatedWork} reviews related work on automated facilitation, conversational agents, and parallel thinking methodologies. Section \ref{sec:methodology} introduces the overall framework of the proposed agent-based facilitation system. 
Section \ref{sec:userStudy} details the implementation and experimental setup, and presents the results of the pilot experiments, along with a discussion of their implications, limitations, and potential directions for future work. 
Finally, Section \ref{sec:conclusion} concludes the paper.

\section{Related Work} \label{sec:relatedWork}

\subsection{Group Decision Support with Automated Facilitation}

Group decision making benefits significantly from facilitation, which can occur during, before, or after group discussions \cite{bostrom1993group,beranek1993facilitation,zarate2013collaborative}. Traditionally, human facilitators have been essential in guiding discussions, requiring expertise in both the subject matter and the language of communication \cite{gimpel_digital_2024}. However, the shortage of skilled facilitators creates a gap that modern automated facilitation systems are increasingly positioned to address \cite{gu_case-based_2021}.
Most existing automated facilitation approaches rely on rule-based and static reasoning systems \cite{gu_case-based_2021}. While these systems provide structure, their lack of adaptability limits their effectiveness in dynamic group interactions. Recently, LLMs have gained prominence as tools for brainstorming and consensus building, offering a dynamic and adaptive alternative. In contrast to rigid systems, LLMs engage in active collaboration with human users, supporting more natural facilitation \cite{noack2025ai,wan_second_mind,Joosten_ideation,bouschery_ai-augmented_2023}.
LLMs have demonstrated promise in several aspects of group decision-making. They facilitate agreement building within groups \cite{Tessler_common_ground,shin2022chatbots}, enhance the quality of discourse \cite{Argyle_democratic_discourse}, and contribute to better decision-making outcomes \cite{ma2024towards}. Additionally, LLMs can foster diversity of opinions \cite{Kim_chi} and improve inclusivity, such as by increasing women's participation in discussions \cite{hadfi_conversational_2023}.

However, prior LLM-based facilitation approaches tend to focus on technical optimization, or surface level coordination \cite{matsumura_2024_ai_facilitation}, overlooking research in other fields that offer structured methods for enhancing group discussion. We argue that effective automated facilitation should be grounded in established methodologies that are designed to support reflective, balanced, and goal-oriented discourse. To this end, this paper explores  the role of LLMS as generalized facilitators, able to perform different tasks and roles, through the lens of the Six Thinking Hats methodology \cite{de2017six}. This structured approach to parallel thinking is well-established in the organizational literature but has not been utilized in automated facilitation. Furthermore, we assess the quality of deliberations from the user’s perspective, a dimension that has been identified as lacking sufficient research \cite{Shortall_2022}.

\subsection{Facilitation with the Six Thinking Hats Method}
The Six Thinking Hats framework is a well-established method for enhancing group decision-making and problem-solving. It involves six metaphorical hats, each representing a distinct mode of thinking. By focusing on one type of thinking at a time, the method reduces confusion and improves efficiency in group discussions \cite{de2017six}.
The Six Thinking Hats methodology has been employed in experimental settings with encouraging results, particularly for creative brainstorming \cite{gocmen_effects_2019,doi:10.1177/2379298116676596}. Previous studies have explored the use of this framework in facilitating human reflection on LLM-generated outputs, where participants evaluate and critique ideas rather than having LLMs actively assume the roles of the six hats \cite{zha2024designing}.

In contrast, our approach assigns each LLM agent the role of one of the Six Thinking Hats to facilitate real-time discussions with human participants. While prior work has explored similar roles for LLMs in agent-to-agent discussions \cite{lu2024llm,mushtaq2025harnessing}, to the best of our knowledge, this study is the first to evaluate their effectiveness in facilitating human group interactions through this method.
The Six Thinking Hats approach can lead to higher-quality argumentation and foster critical thinking \cite{ekahitanond2018adopting}. However, it may also increase the cognitive load for participants, requiring careful consideration of its implementation \cite{ma2024strengthening}. By leveraging LLMs to assume these roles, we aim to mitigate the cognitive burden on human participants while enhancing the overall quality and depth of discussions.
Our study is the first to assess the impact of using LLMs to actively embody the Six Thinking Hats roles in real-time group discussions, providing new insights into their potential for automated facilitation and collaborative decision-making.

\section{Methodology}
\label{sec:methodology}

\textbf{\textit{Overall Implementation.}}
To support the PTFA framework, we developed an online discussion platform based on the Discourse forum system\footnote{https://www.discourse.org/}, enhanced with large language model (LLM) integration. The platform enables structured, real-time text-based discussions, with LLM agents acting as facilitators according to the Six Thinking Hats methodology. The system architecture consists of a web-based interface for user participation, a backend database to store user-generated content and metadata, and an integration with the OpenAI Assistants API\footnote{https://openai.com/} for dynamic response generation. Communication between frontend and backend components is handled using WebSockets to ensure real-time responsiveness.

To manage discussions securely and effectively, each participant is assigned a unique anonymous identifier, and all discussion data is stored in a PostgreSQL database. The platform employs HTTPS encryption for secure communication and role-based access control to restrict system access. In addition, third-party plugins are used to support logging, moderation, and analytics, providing additional functionality for monitoring and evaluating discussion quality. The LLM agents are triggered based on predefined time intervals or inactivity detection, ensuring responsive but non-intrusive facilitation throughout the conversation.

\begin{figure}
  \includegraphics[width=0.9\textwidth]{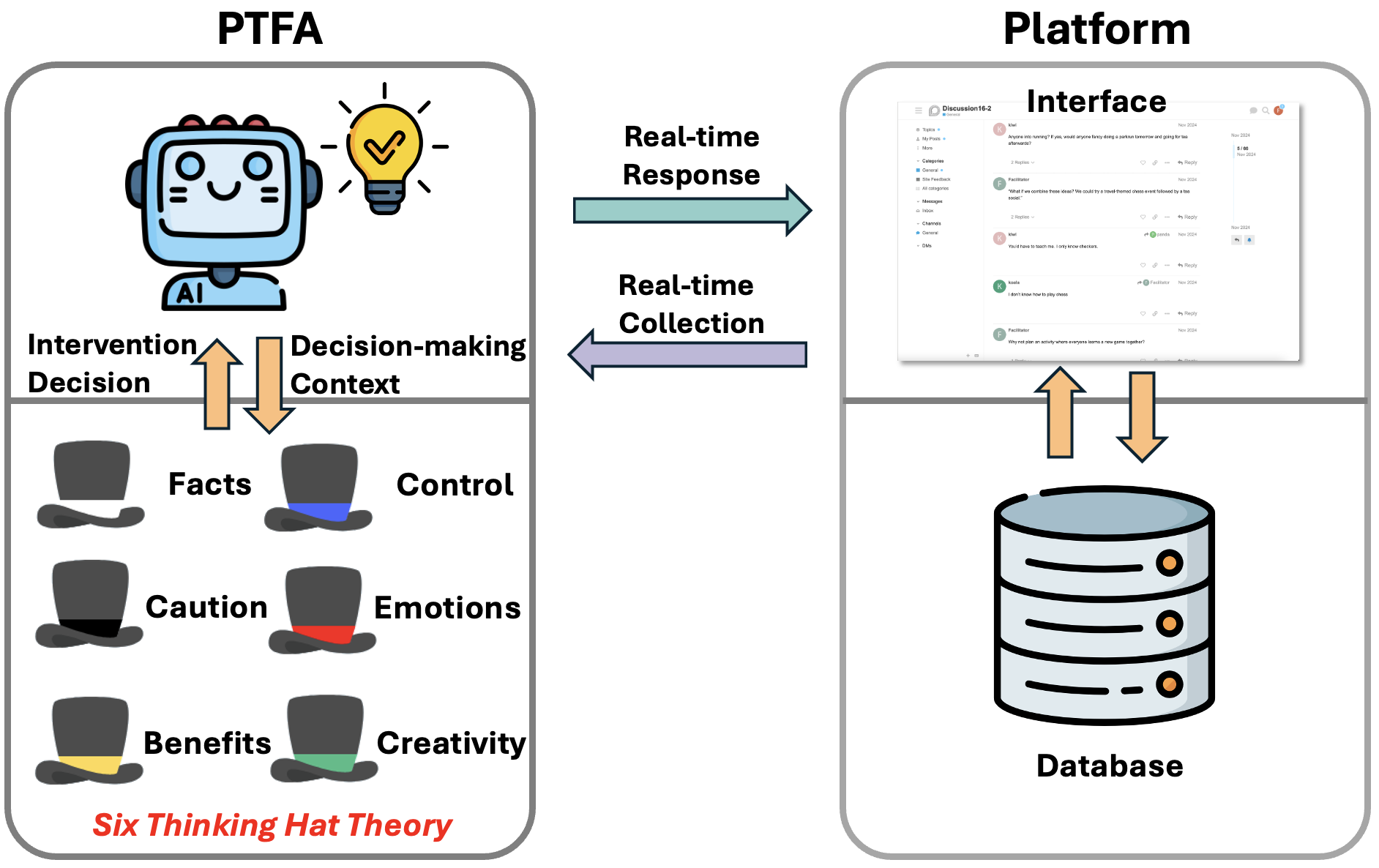}
      \caption{The framework of PTFA.}
  \label{fig:Framework}
\end{figure}

\textbf{\textit{Parallel Thinking-based Facilitation Agent Implementation.}}  The facilitation agents in the system are implemented using OpenAI's ChatGPT assistant API, which enables real-time responses to guide structured discussions based on the Six Thinking Hats framework (shown in Fig.\ref{fig:Framework}). Each thinking hat is represented by a distinct LLM agent that provides contextually appropriate responses. 

The implementation of LLM-based facilitation agents requires careful prompt engineering at both a macro and micro level. At a macro level, prompts should be designed to provide comprehensive context by outlining the discussion background, defining the agent's role, and specifying the overall objectives. This includes instructing the agent to guide participants towards consensus, foster balanced discussions, and adhere to structured communication principles. At the micro level, prompts should incorporate specific instructions that allow the agent to dynamically assess the current state of the conversation and intervene appropriately. These prompts should include mechanisms to determine when intervention is necessary, such as identifying signs of topic drift, unresolved conflicts, or opportunities for further elaboration. The prompts also define response styles, ensuring that the agent's interventions are concise, relevant, and aligned with the discussion phase. The prompt engineering for each of the six hat agents is detailed below:

\begin{itemize}
    \item \textbf{White Hat (Facts and Information):} This agent embodies objective thinking. Its prompt instructs it to focus solely on data and facts without interpretation, grounding the conversation in reality. It is designed to intervene when the discussion requires objective information or when participants make unsupported statements.
    \textit{An example of a situational prompt is: "Could you clarify the exact figures or facts related to this issue? Here's what we know so far: [insert relevant data]."}

    \item \textbf{Red Hat (Emotions and Intuition):} Representing feelings and gut reactions, this agent allows for the expression of emotion without justification. The primary prompt directs it to provide a short, intuitive perspective when the conversation touches on personal feelings.
    \textit{A situational prompt for this agent is: "This feels like an emotional moment. How are we feeling about this issue right now?"}

    \item \textbf{Black Hat (Critical Thinking and Risk):} This agent provides a critical, cautious perspective. Its core instruction is to identify potential risks, weaknesses, and flaws that the group might be overlooking, ensuring negative outcomes are considered.
    \textit{An example of its situational prompt is: "Have we considered the potential downsides? Here's a risk we might be overlooking: [insert risk]."}

    \item \textbf{Yellow Hat (Optimism and Benefits):} The Yellow Hat agent focuses on optimism and positive thinking. It is prompted to offer a constructive view by highlighting the benefits and potential advantages of ideas, especially when the discussion needs a boost of optimism.
    \textit{A corresponding situational prompt is: "Looking at the bright side, this idea offers some exciting opportunities we shouldn’t overlook."}

    \item \textbf{Green Hat (Creativity and Alternatives):} This agent encourages creative and innovative thinking. It is prompted to intervene with new ideas or alternative approaches when the conversation becomes stalled or could benefit from novel solutions.
    \textit{An example of a creative situational prompt is: "What if we approached this from a different angle? Here’s an idea to consider: [insert new idea]."}

    \item \textbf{Blue Hat (Process and Control):} The Blue Hat agent manages the thinking process itself. Its instructions are to provide structure and direction, organize the conversation, and ensure the discussion stays on track and that all perspectives are considered.
    \textit{A situational prompt for managing the process is: "It seems like we’re getting off track. Maybe we should focus on this key point: [insert key point]."}
\end{itemize}

To minimize unnecessary interruptions and maintain a smooth discussion flow, a special mechanism is implemented in which the LLM outputs 'Good' when it determines that the current conversation is progressing well without the need for further intervention. These outputs are automatically filtered and hidden from the discussion interface to avoid disrupting the participants. This approach ensures that the system provides support only when necessary, enhancing the overall user experience by reducing distractions while maintaining the integrity of the discussion. 

Prompt engineering for these LLM-based facilitation agents is designed to ensure that each agent's responses align with its designated role while maintaining clarity and conciseness. The primary prompts provide general guidance based on the respective thinking mode, whereas situational prompts encourage specific interventions tailored to the evolving discussion context. For instance, the White Hat prompts focus on factual accuracy and verification, while the Green Hat encourages brainstorming new ideas. The agents are triggered based on predefined timing intervals. Timing intervals are set based on the progression of the discussion phase, with interventions to ensure continuous engagement without overwhelming the participants. If the system detects a lack of diversity in input or prolonged inactivity, the agents are prompted to re-engage users with targeted suggestions or requests for further elaboration.

\section{User Study}
\label{sec:userStudy}
\subsection{Participants \& Study Setting}
\label{sec:userStudy:participantsStudySetting}

\begin{figure}[htb]
\includegraphics[width=0.9\textwidth]{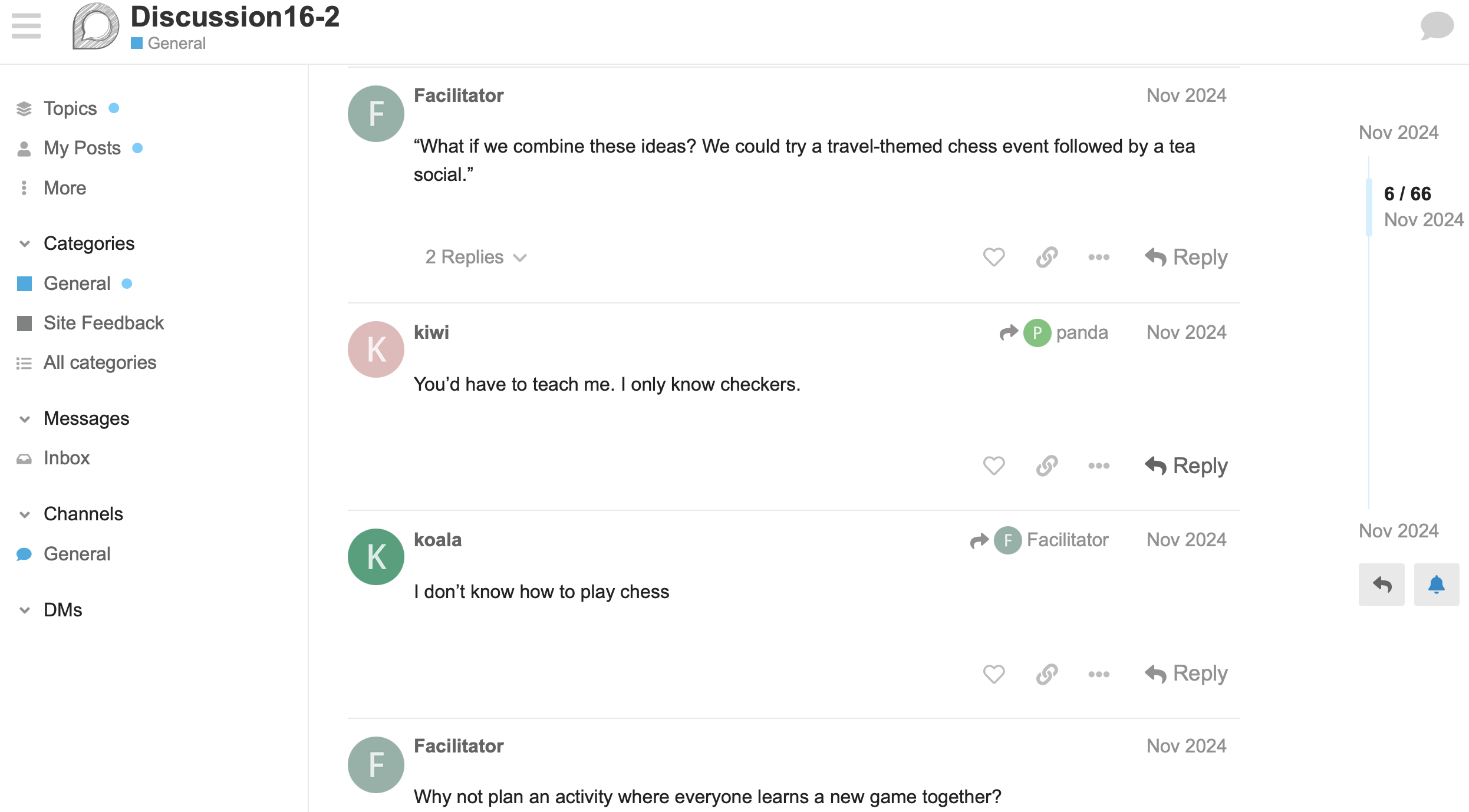}
  \caption{Screenshot of online discussion platform}
  \label{fig:system}
\end{figure}

To evaluate the efficacy of the proposed PTFA in supporting online consensus building, we held text-based real-world online discussions.
Before embarking on this study, we secured ethical approval from the Faculty Ethics Committee at University of Southampton, guaranteeing adherence to established guidelines for ethical conduct in research involving human subjects.
A total of 48 participants were recruited through lecture announcements with faculty permission, email invitations via department mailing lists, and posters placed in university common areas.
16 discussion groups were formed and each discussion group consisted of 3 participants, who were randomly selected while ensuring a balance in gender and English proficiency levels in each group.
Specifically, each group included at least two different genders and participants with at least two different levels of English proficiency. 
The experiments were conducted in university tutorial rooms equipped with laptops that connected to the online discussion platform we developed.
Figure \ref{fig:system} shows the screenshot of the online discussion platform.
Each discussion group participated in two online discussions, each featuring a different combination of discussion topics and facilitation models and followed by a short facilitator evaluation survey. The purpose of the survey was to evaluate the participants' perception of the facilitator's effectiveness, overall user experience and the degree to which consensus was achieved. By comparing the responses between the facilitator models, we aim to evaluate the practical impact of PTFA on group discussion dynamics and outcomes. Through this process we identify areas of improvement for LLM-assisted facilitation. 

Each discussion session lasted for 20 minutes.
To alleviate barriers to participation, two topics related to daily life were selected as discussion subjects. 
Topic0 was "Please decide one activity that you would like to do together" and Topic1 was "Please decide one film that you would like to watch together".
Additionally, two facilitation models were designed for comparison.
Facilitation Model 0 represented a traditional facilitation model in which three facilitation messages were generated, each after a certain amount of elapsed time:
\begin{itemize}
    \item ``Hi all, our goal today is to reach a consensus on the question posed at the end of the discussion. Please start by generating ideas.'' was posted at the beginning of the discussion,
    \item  ``You have already discussed it for 10 mins. This is a good time for you to reconsider the ideas that you have already had.'' was posted 10 minutes after the discussion began, and
    \item ``There are only 3 minutes left, if you haven’t reached a consensus yet, please make a decision as soon as possible.'' was posted 17 minutes after the discussion started. 
\end{itemize}
In contrast, Facilitation Model 1 was developed using the proposed PTFA approach, introducing an alternative facilitation method for comparison.
Intervention decisions are made at 30-second intervals, taking group size into account.

\subsection{Results}
\label{sec:results}

\begin{figure}[htb]
    \centering
    \hfill
    \begin{subfigure}[t]{0.32\textwidth}
        \centering
        \includegraphics[width=\textwidth]{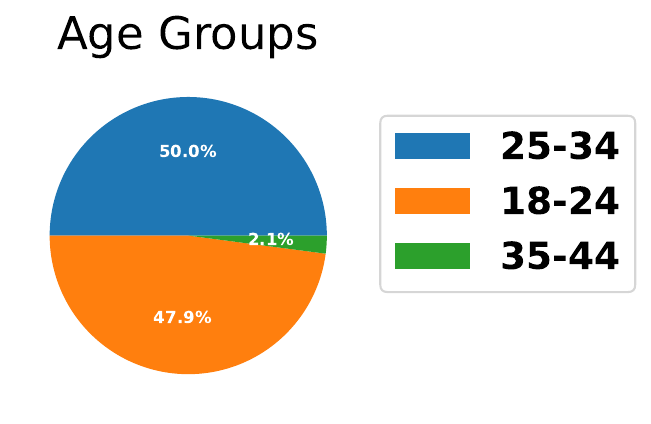}
        \caption{The share of the participants by their ages.}
        \label{fig:participants:age}
    \end{subfigure}
    \hfill
    \begin{subfigure}[t]{0.45\textwidth}
        \centering
        \includegraphics[width=\textwidth]{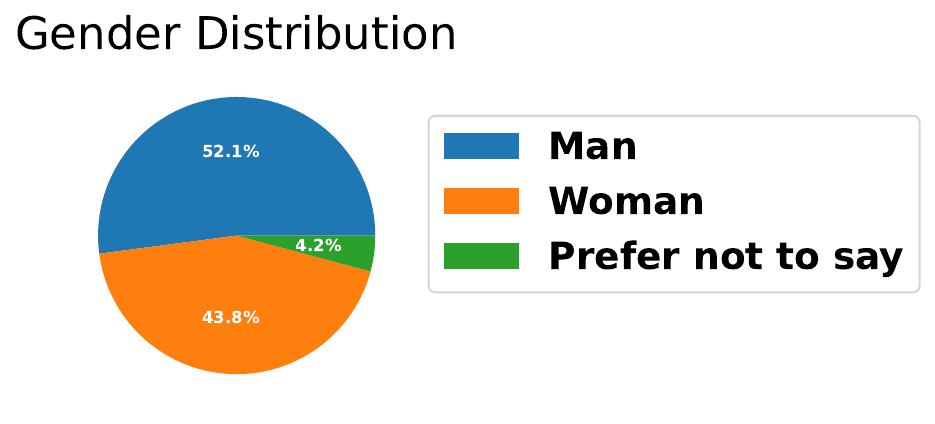}
        \caption{The share of the participants by their genders.}
        \label{fig:pariticpants:gender}
    \end{subfigure}\hfill\\
    \begin{subfigure}[t]{0.5\textwidth}
        \centering
        \includegraphics[width=\textwidth]{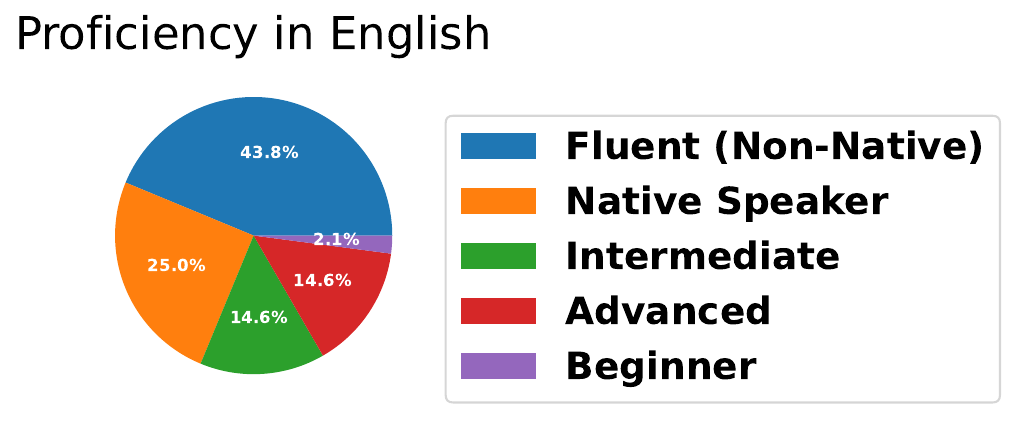}
        \caption{The share of the participants by their proficiencies in English.}
        \label{fig:participants:englishProficiency}
    \end{subfigure}
    \caption{The characteristics of the participants in terms of age, gender and  English proficiency.}

    \label{fig:participants}
\end{figure}

Our study consisted of 48 participants, with 3 in each of the 16 groups. Each group participated in two discussion giving a dataset of 32 discussions containing 16,656 words in 1,669 posts, of which the facilitator contributed 3,459 words in 217 posts. Each post has a corresponding timestamp and, for the LLM-generated posts, which hat was used to generate the text. Participants were almost entirely aged between 18 and 34, with a near even split between males and females. Most participants (68.8\%) considered themselves a fluent or native speaker of english, with only 2.1\% considering themselves a beginner, the least proficient category. A full breakdown of the participant demographics is presented in Figure \ref{fig:participants}.

Analysis of the discussion transcripts reveals that while there are promising indications of the efficacy of PTFA-based consensus building, there are also challenges in utilizing an LLM to make timely, appropriate contributions that consistently advance the discussion in a productive manner. The post-discussion surveys support this observation. The survey responses indicate that participants had more mixed feelings about the performance of the PTFA facilitator on a number of discussion attributes, including the extent to which the facilitator helped consensus decision making (see Figure~\ref{fig:surveyResults:facilitatorRating}), when compared to a baseline facilitation approach. Below, we first present the survey results that provide an overview of participant experiences. We then examine the types of constructive conversation patterns observed in the discussions before analyzing the problematic timing and formulation of comments that likely contributed to the lower satisfaction scores.

\subsubsection{Survey Responses}

The complementing surveys (see also Section~\ref{sec:userStudy:participantsStudySetting}) were designed to evaluate the level of support and capabilities the PTFA approach already provides, which, while highlighting a general positive view, emphasize how much problematic LLM patterns (see Section~\ref{sec:results:problematicFacilitationPatterns}) can dampen the facilitation experience and thus the overall experience and outcomes.

\begin{figure}[htb]
    \centering
    \hfill
    \begin{subfigure}[t]{0.3\textwidth}
        \centering
        \includegraphics[width=\textwidth]{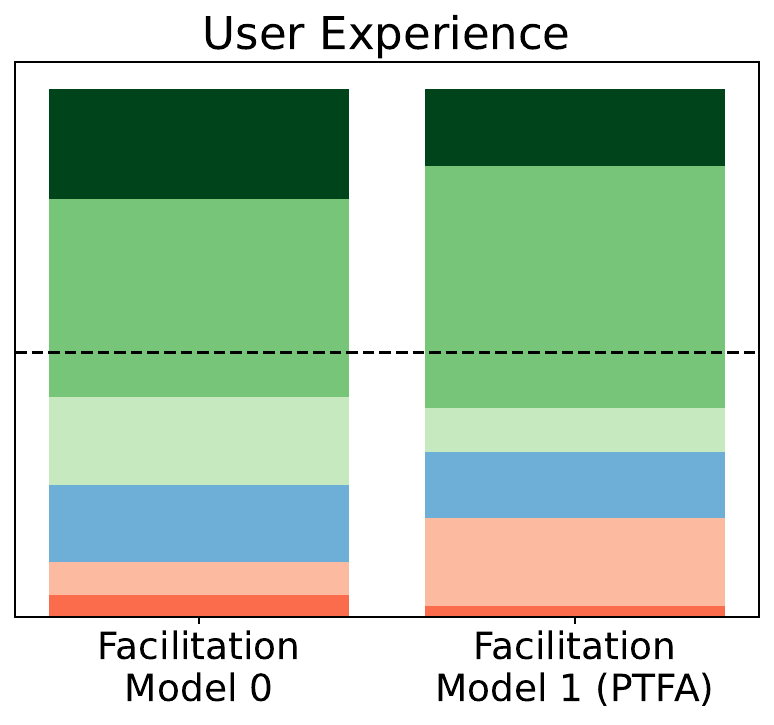}
        \caption{Survey responses for the question `How would you rate the user experience of the platform?' rated into 7 possible responses as indicated in the legend in Figure~\ref{fig:surveyResults:legendSatisfaction}.}
        \label{fig:surveyResults:userExperience}
    \end{subfigure}
    \hfill
    \begin{subfigure}[t]{0.3\textwidth}
        \centering
        \includegraphics[width=\textwidth]{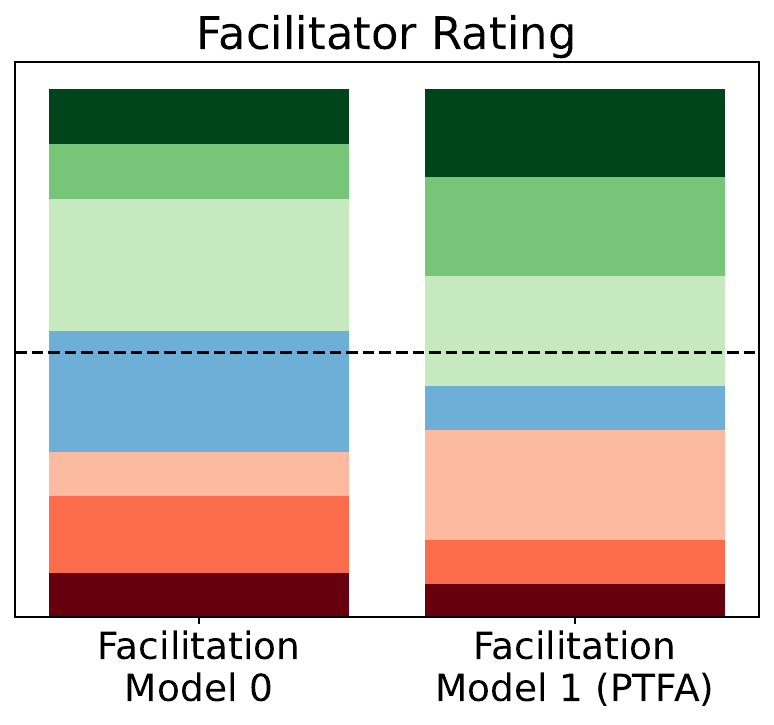}
        \caption{Survey responses for the question `How would you rate the extent to which the facilitator in the discussion helped consensus decision-making?' rated into 7 possible responses as indicated in the legend in Figure~\ref{fig:surveyResults:legendAgreement}.}
        \label{fig:surveyResults:facilitatorRating}
    \end{subfigure}
    \hfill
    \begin{subfigure}[t]{0.3\textwidth}
        \centering
        \includegraphics[width=\textwidth]{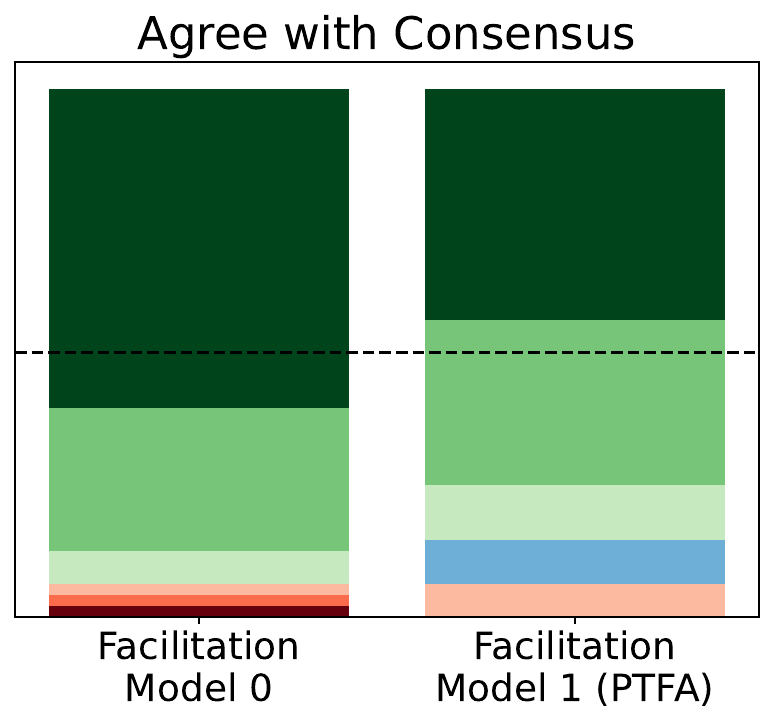}
        \caption{Survey responses for the question `Do you agree with the consensus reached in this discussion?' rated into 7 possible responses as indicated in the legend in Figure~\ref{fig:surveyResults:legendAgreement}.}
        \label{fig:surveyResults:consensusAgreement}
    \end{subfigure}\hfill\\
    \begin{subfigure}[t]{0.4\textwidth}
        \centering
        \includegraphics[width=\textwidth]{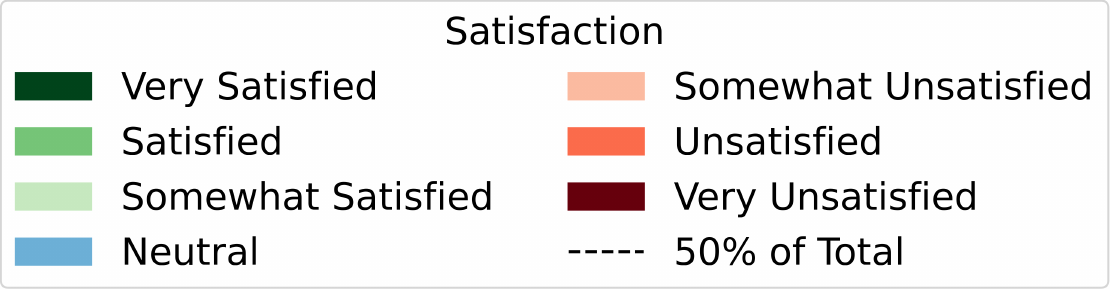}
        \caption{The legend for Figure~\ref{fig:surveyResults:userExperience}}
        \label{fig:surveyResults:legendSatisfaction}
    \end{subfigure}
    \hspace{1.0cm}
    \begin{subfigure}[t]{0.37\textwidth}
        \centering
        \includegraphics[width=\textwidth]{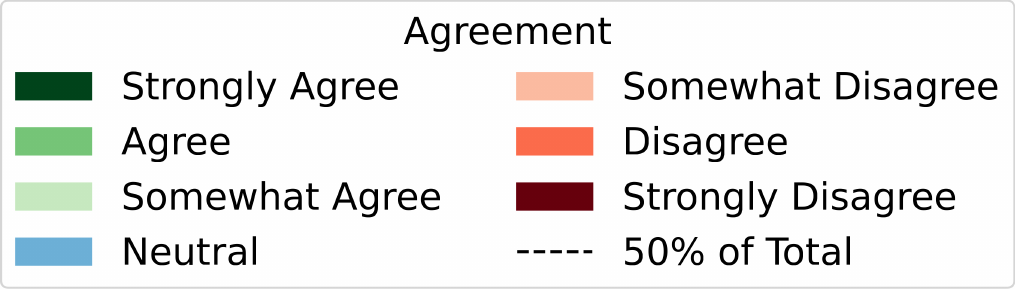}
        \caption{The legend for Figure~\ref{fig:surveyResults:facilitatorRating} and Figure~\ref{fig:surveyResults:consensusAgreement}}
        \label{fig:surveyResults:legendAgreement}
    \end{subfigure}
    \caption{Survey responses for the questions around the rating of the facilitator models and the agreement with the resulting consensus.
        The legend in Figure~\ref{fig:surveyResults:legendAgreement} shows the possible responses used for the questions in Figure~\ref{fig:surveyResults:facilitatorRating} and Figure~\ref{fig:surveyResults:consensusAgreement}.
            The 50\% line divides the bars at 50\% of participants.}
    \label{fig:surveyResults:facilitatorRatingAndConsensusAgreement}
\end{figure}

Generally, the user experience with either facilitation model has been positive in terms of user experience (see Figure~\ref{fig:surveyResults:userExperience}) and reached consensus (see Figure~\ref{fig:surveyResults:consensusAgreement}).
However, despite positive contributions by the PTFA-based facilitator (Facilitation Model 1) towards being an effective and practical consensus building facilitator (see Section~\ref{sec:results:constructiveFacilitationPatterns}), negative effects (see Section~\ref{sec:results:problematicFacilitationPatterns}) lead to mixed facilitator ratings.
We observe this in the response to the question `How would you rate the extent to which the facilitator in the discussion helped consensus decision-making?' to which the participant could indicate their agreement with one of the statements: `Strongly Agree', `Agree', `Somewhat Agree', `Neutral', `Somewhat Disagree', `Disagree' and `Strongly Disagree'.
While participants reacted more positively to the PTFA-based facilitator - being slightly above 50\% - in comparison to Facilitation Model 0 - being slightly below 50\% - as Figure~\ref{fig:surveyResults:facilitatorRating} shows, overall, the rating of both is mixed.

The mixed performance of the PTFA facilitator also affects the user experience and their agreement with the consensus.
As Figure~\ref{fig:surveyResults:userExperience} highlights, while the user experience is still positive, it is slightly dampened by some of the problematic LLM patterns (see Section~\ref{sec:results:problematicFacilitationPatterns}).
The user experience assessment is based on the question `How would you rate the user experience of the platform?' to which the participation could indicate their satisfaction with 7 the responses: `Very Satisfied', `Satisfied', `Somewhat Satisfied', `Neutral', `Somewhat Unsatisfied', `Unsatisfied' and `Very Unsatisfied'.
Moreover, the PTFA facilitator also slightly dampens the agreement with the consensus, especially due to disrupting the consensus forming with the LLM's difficulty to keep to specified phases (see~\ref{par:results:problematicFacilitationPatterns:poorPhaseManagement} in Section~\ref{sec:results:problematicFacilitationPatterns}).
This consensus agreement assessment is based on the question `Do you agree with the consensus reached in this discussion?' to which the participants could indicate their agreement with the same 7 responses as above.

\subsubsection{Constructive Facilitation Patterns}
\label{sec:results:constructiveFacilitationPatterns}

\begin{figure}[t]
\begin{subfigure}{0.48\textwidth}
\includegraphics[width=\textwidth]{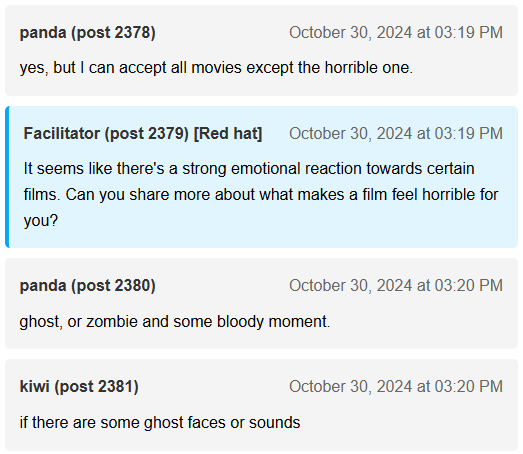}
\caption{The red hat externalizing its analysis of the participants comment.}
\label{fig:discussion:red_hat}
\end{subfigure}
\hfill
\begin{subfigure}{0.48\textwidth}
\includegraphics[width=\textwidth]{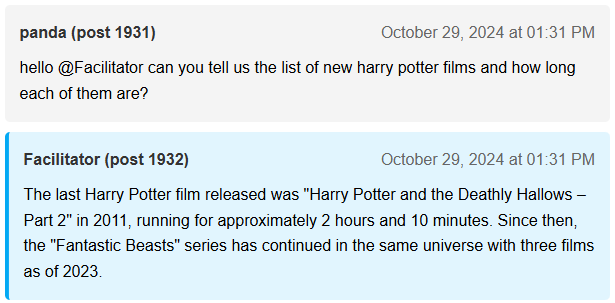}
\caption{The white hat responding with accurate factual information to a question posed by a participant.}
\label{fig:discussion:white_hat}
\end{subfigure}
\caption{Examples of constructive facilitator contributions}
\label{fig:discussion:constructive_examples_1}
\end{figure}

\begin{figure}[t]
\begin{subfigure}{0.48\textwidth}
\includegraphics[width=\textwidth]{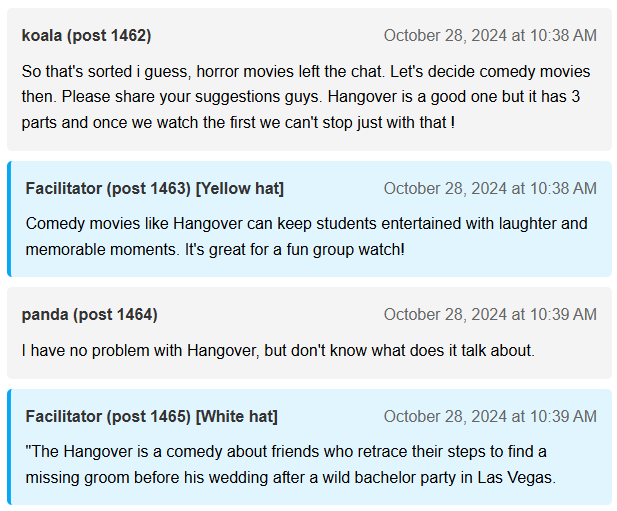}
\caption{The yellow hat highlighting positives of a movie. The white hat providing accurate factual information about a movie in response to a participant expressing their unfamiliarity with it.}
\label{fig:discussion:yellow_white_hat}
\end{subfigure}
\hfill
\begin{subfigure}{0.48\textwidth}
\includegraphics[width=\textwidth]{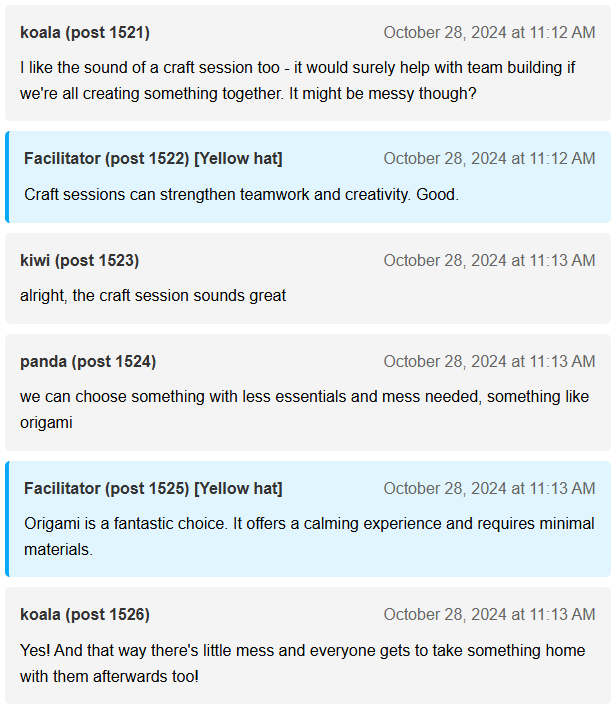}
\caption{The yellow hat highlighting positive aspects of ideas that resonate with the participants.}
\label{fig:discussion:yellow_hat}
\end{subfigure}
\caption{Examples of constructive facilitator contributions}
\label{fig:discussion:constructive_examples_2}
\end{figure}

We observed cases in which the PTFA facilitator made productive contributions to the discussion that were distinct from the types of contributions provided by the human participants. We identify four types of useful intervention, each of which was contributed by a particular thinking hat. In the following, we describe these useful interventions. Example discussion excerpts of positive contributions are presented in Figure \ref{fig:discussion:constructive_examples_1} and Figure \ref{fig:discussion:constructive_examples_2}.

\textbf{\textit{Highlighting and exploring emotional responses of participants (Red Hat).}}
The human participants mostly gave shallow opinions about suggestions; they typically said little more than whether or not they liked an idea. By asking focused questions about the participants' emotional reactions and feelings, the PTFA facilitator helped them make better informed decisions about their existing ideas and generate new ones better aligned with their preferences.

\textbf{\textit{Identifying positive/negative qualities of suggestions that meet the participants needs (Yellow/Black Hat).}} 
Identifying strengths and weaknesses of suggestions, the PTFA facilitator provided a deeper analysis of proposals than that typically provided by the human participants independently. These comments highlighted particular features of ideas that aligned or conflicted with participants' expressed interests. Contributions of this type usually expanded on a positive or negative comment made by a participant.

\textbf{\textit{Providing factual information about the discussion topic (White Hat).}} 
The LLM is able to generate relevant facts learned from its training data. It did this during discussions, either when participants expressed that they did not know some factual information (for example, the runtime of a film) or explicitly asked the facilitator for such information. While it is possible for LLMs to hallucinate details, almost all the facts it generated were correct. One exception was a comment that a theater production would open in the future, even though it had actually already started. Inaccuracies like this are due to the cut-off date of the data used to train the model rather than the hallucination of facts.

\textbf{\textit{Introducing new suggestions to the discussion (Green Hat).}} 
The PTFA facilitator regularly provided or extended suggestions. These suggestions were often well-received by participants and then selected as the consensus decision. The LLM has a tendency to combine or extend prior suggestions, particularly when generating ideas for a shared activity. While these extensions were also sometimes well received, on other occasions, the participants joked about how many components the facilitator was adding to the activities.

\subsubsection{Problematic Facilitation Patterns}
\label{sec:results:problematicFacilitationPatterns}

\begin{figure}[t]
\begin{subfigure}{0.48\textwidth}
\includegraphics[width=\textwidth]{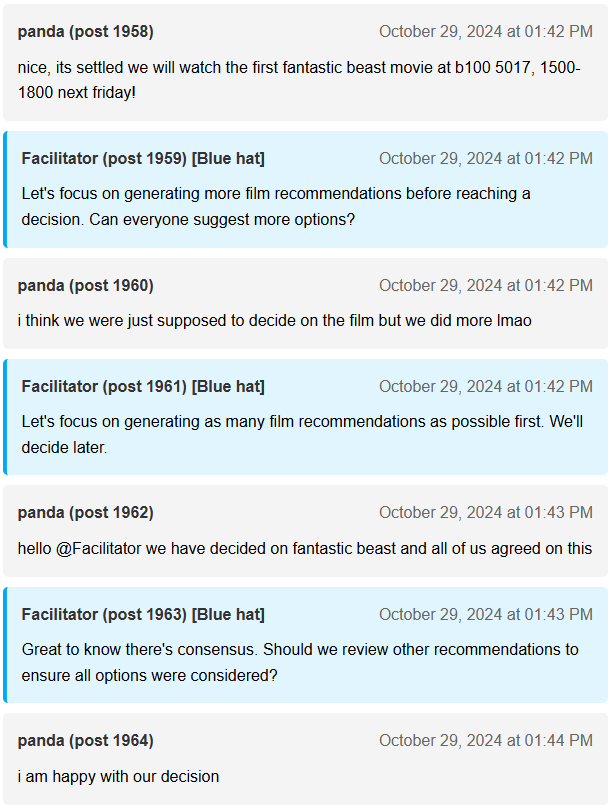}
\caption{Asking for idea suggestions repeatedly and at an inappropriate time. These comments begin 4 minutes before the end of the 20 minute discussion period.}
\label{fig:discussion:many_recommendations}
\end{subfigure}
\hfill
\begin{subfigure}{0.48\textwidth}
\includegraphics[width=\textwidth]{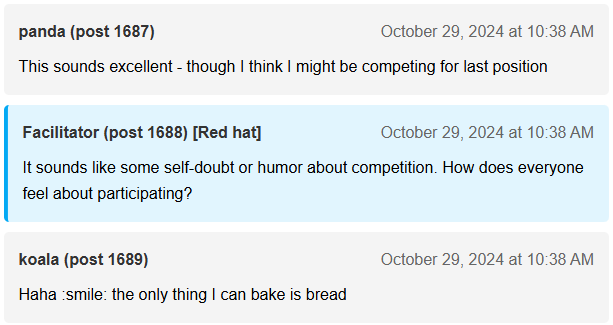}
\caption{The red hat externalizing its analysis of a participant's comment.}
\label{fig:discussion:red_unnatural}
\end{subfigure}
\caption{Examples of problematic facilitator behaviors.}
\label{fig:discussion:problematic_examples}
\end{figure}

While the PTFA facilitator is capable of making useful contributions, these are offset by the many comments that were unhelpful or even counterproductive. We observed two main categories of issues: poor management of the discussion phase and inconsistent response timing. Example discussion excerpts of problematic contributions are presented in Figure \ref{fig:discussion:problematic_examples}.

\textbf{\textit{Poor Phase Management.}} 
\label{par:results:problematicFacilitationPatterns:poorPhaseManagement}
The most significant of these issues is in scheduling the phases of the consensus building discussion. The LLM is prompted to facilitate a divergent first discussion stage, encouraging idea generation, followed by a convergent second stage of consensus formation.
In practice, the facilitator stayed in the first stage throughout the discussions, with the green and blue hat personas consistently asking for further suggestions until the end of the 20-minute period, including when participants were trying to form a consensus or explicitly stated that they had reached a consensus they were happy with.
It also asked the participants to generate new suggestions while they were already in the process of sharing new suggestions.
Often, these inappropriate requests for more ideas were ignored by the participants. However, in one discussion, a consensus was reached until the facilitator intervened with more suggestions. The new ideas split the participants' preferences without enough time for them to discuss and agree on a new unanimous decision.
Hence, to build an effective facilitator based on LLMs, it is necessary to allow the facilitator to be aware of phases and times and fine-tune the facilitator to judge the state of the discussion from the participants' perspective.

\textbf{\textit{Inconsistent response timing.}} 
Another issue is that the PTFA facilitator sometimes did not respond for long periods of time.
In one instance, it makes its first contribution 8 minutes into the session, with a comment ``starting'' the discussion - despite the participants already having generated many ideas.
Similarly, it sometimes ignores questions explicitly asked of it.
On the other hand, the PTFA facilitator may intervene too much, responding after every individual comment without allowing the participants to develop their ideas.
These repeated messages are typically those asking for more suggestions, as described above.
Hence, it is necessary to be aware that LLMs can interact differently in seemingly similar situations and fine-tune the LLM's intervention rate based on the state of the discussion, with respect to the content, and its previous interventions.

\subsection{Discussion}
This study provides insight into both the potential and limitations of LLM-augmented discussions. Analysis of the 32 facilitated discussions shows that in a PTFA-based framework, LLMs can make valuable contributions that are otherwise neglected by human participants, particularly through idea generation, emotional probing, and deeper analysis of positive and negative qualities of ideas. These findings demonstrate the potential utility of LLMs in discussion facilitation, offering examples of how a PTFA could automate skilled facilitation to improve outcomes of consensus building.
Despite this, the survey results reveal that users had more mixed opinions of the PTFA facilitator when compared to a well-received baseline facilitator on several discussion attributes. This is supported by the qualitative analysis of discussion transcripts, in which we identified many instances of the PTFA facilitator making unhelpful or inappropriate contributions to the discussion.

The timing of LLM interventions proved particularly problematic. Often, these contributions are inappropriate because they come at the wrong time in the discussion. For example, when it asks for suggestions at the end of the discussion or gives an emotional analysis of a user's non-emotional comment. On the other hand, the PTFA facilitator is also sometimes absent from the discussion and misses moments when an intervention could prove useful, such as when asked a direct question by a participant or when the participants need to converge on a consensus decision. This highlights the need to investigate how to control LLM contributions in multi-participant conversations, an area less studied than back-and-forth interactions with a single other participant.

\section{Conclusions} \label{sec:conclusion}

This paper explored the potential of AI agents in supporting consensus building processes in a natural conversation scenario. 
We introduced the Parallel Thinking-based Facilitation Agent (PTFA) that utilized parallel thinking-based strategies aligned with general-purpose LLMs. 
A real-world pilot user study, including 48 participants, was conducted, demonstrating the PTFA's ability while also identifying a number of remaining challenges.
We not only highlight insights on how LLMs are suitable but also note some limitations and open challenges, thus providing directions for utilizing LLMs effectively for facilitation agents.
Moreover, based on the user study, we constructed a novel dataset rich in conversational content among participants and between participants and PTFA.
Future work will focus on addressing the open challenges identified in the user study.

\begin{credits}
\subsubsection{\ackname} This work was supported by JST CREST JPMJCR20D1, and JSPS KAKENHI Grant Number
JP24K20900. 
Additional support for this work was provided by the UK Engineering and Physical Sciences Research Council (EPSRC) through a Turing AI Fellowship (EP/V022067/1) on Citizen-Centric AI Systems (\url{https://ccais.ac.uk/}) and through the FEVER Programme Grant (EP/W005883/1). 
\end{credits}

%
%
%
\bibliographystyle{splncs04}
\bibliography{mybibliography}

\begin{thebibliography}{10}
\providecommand{\url}[1]{\texttt{#1}}
\providecommand{\urlprefix}{URL }
\providecommand{\doi}[1]{https://doi.org/#1}

\bibitem{Argyle_democratic_discourse}
Argyle, L.P., Bail, C.A., Busby, E.C., Gubler, J.R., Howe, T., Rytting, C., Sorensen, T., Wingate, D.: Leveraging ai for democratic discourse: Chat interventions can improve online political conversations at scale. Proceedings of the National Academy of Sciences  \textbf{120}(41),  e2311627120 (2023). \doi{10.1073/pnas.2311627120}

\bibitem{joshua_network_2017}
Becker, J., Brackbill, D., Centola, D.: Network dynamics of social influence in the wisdom of crowds. Proceedings of the National Academy of Sciences  \textbf{114}(26),  E5070--E5076 (2017). \doi{10.1073/pnas.1615978114}

\bibitem{beranek1993facilitation}
Beranek, P.M., Beise, C.M., Niederman, F.: Facilitation and group support systems. In: [1993] Proceedings of the Twenty-sixth Hawaii International Conference on System Sciences. vol.~4, pp. 199--207. IEEE (1993)

\bibitem{bostrom1993group}
Bostrom, R.P., Anson, R., Clawson, V.K.: Group facilitation and group support systems. Group support systems: New perspectives  \textbf{8},  146--168 (1993)

\bibitem{bouschery_ai-augmented_2023}
Bouschery, S.G., Blazevic, V., Piller, F.T.: {AI}-{Augmented} {Creativity}: {Overcoming} the {Productivity} {Loss} in {Brainstorming} {Groups}. Academy of Management Proceedings  \textbf{2023}(1),  11938 (Aug 2023). \doi{10.5465/AMPROC.2023.11938abstract}, publisher: Academy of Management

\bibitem{de2017six}
De~Bono, E.: Six Thinking Hats: The multi-million bestselling guide to running better meetings and making faster decisions. Penguin uk (2017)

\bibitem{ekahitanond2018adopting}
Ekahitanond, V.: Adopting the six thinking hats to develop critical thinking abilities through line. Australian Educational Computing  \textbf{33}(1),  1--16 (2018)

\bibitem{doi:10.1177/2379298116676596}
Fender, C.M., Stickney, L.T.: When two heads aren’t better than one: Conformity in a group activity. Management Teaching Review  \textbf{2}(1),  35--46 (2017). \doi{10.1177/2379298116676596}

\bibitem{gimpel_digital_2024}
Gimpel, H., Lahmer, S., Wöhl, M., Graf-Drasch, V.: Digital {Facilitation} of {Group} {Work} to {Gain} {Predictable} {Performance}. Group Decision and Negotiation  \textbf{33}(1),  113--145 (Feb 2024). \doi{10.1007/s10726-023-09856-8}

\bibitem{gu_case-based_2021}
Gu, W., Moustafa, A., Ito, T., Zhang, M., Yang, C.: A {Case}-based {Reasoning} {Approach} for {Supporting} {Facilitation} in {Online} {Discussions}. Group Decision and Negotiation  \textbf{30}(3),  719--742 (Jun 2021). \doi{10.1007/s10726-021-09731-4}

\bibitem{gocmen_effects_2019}
Göçmen, O., Coşkun, H.: The effects of the six thinking hats and speed on creativity in brainstorming. Thinking Skills and Creativity  \textbf{31},  284--295 (2019). \doi{https://doi.org/10.1016/j.tsc.2019.02.006}

\bibitem{hadfi_conversational_2023}
Hadfi, R., Okuhara, S., Haqbeen, J., Sahab, S., Ohnuma, S., Ito, T.: Conversational agents enhance women's contribution in online debates. Scientific Reports  \textbf{13}(1),  14534 (Sep 2023). \doi{10.1038/s41598-023-41703-3}

\bibitem{ito_an_2022}
Ito, T., Hadfi, R., Suzuki, S.: An {Agent} that {Facilitates} {Crowd} {Discussion}. Group Decision and Negotiation  \textbf{31}(3),  621--647 (Jun 2022). \doi{10.1007/s10726-021-09765-8}

\bibitem{Joosten_ideation}
Joosten, J., Bilgram, V., Hahn, A., Totzek, D.: Comparing the ideation quality of humans with generative artificial intelligence. IEEE Engineering Management Review  \textbf{52}(2),  153--164 (2024). \doi{10.1109/EMR.2024.3353338}

\bibitem{Kim_chi}
Kim, S., Eun, J., Oh, C., Suh, B., Lee, J.: Bot in the bunch: Facilitating group chat discussion by improving efficiency and participation with a chatbot. In: Proceedings of the 2020 CHI Conference on Human Factors in Computing Systems. p. 1–13. Association for Computing Machinery (2020). \doi{10.1145/3313831.3376785}

\bibitem{lu2024llm}
Lu, L.C., Chen, S.J., Pai, T.M., Yu, C.H., Lee, H., Sun, S.H.: Llm discussion: Enhancing the creativity of large language models via discussion framework and role-play (2024), \url{https://arxiv.org/abs/2405.06373}

\bibitem{ma2024towards}
Ma, S., Chen, Q., Wang, X., Zheng, C., Peng, Z., Yin, M., Ma, X.: Towards human-ai deliberation: Design and evaluation of llm-empowered deliberative ai for ai-assisted decision-making (2024), \url{https://arxiv.org/abs/2403.16812}

\bibitem{ma2024strengthening}
Ma, Z.Q., Kong, L.Y., Tu, Y.F., Hwang, G.J., Lyu, Z.Y.: Strengthening collaborative argumentation with interactive guidance: a dialogic peer feedback approach based on the six thinking hats strategy. Interactive Learning Environments  \textbf{33}(1),  1--25 (2024)

\bibitem{matsumura_2024_ai_facilitation}
Matsumura, T., Kato, T., Asa, Y., Esaki, K., Mine, R., Mizuno, H.: Ai-facilitation for consensus-building by virtual discussion using large language models. In: PRICAI 2024: Trends in Artificial Intelligence. pp. 206--219. Springer Nature Singapore (2025)

\bibitem{mushtaq2025harnessing}
Mushtaq, A., Naeem, M.R., Ghaznavi, I., Taj, M.I., Hashmi, I., Qadir, J.: Harnessing multi-agent llms for complex engineering problem-solving: A framework for senior design projects (2025), \url{https://arxiv.org/abs/2501.01205}

\bibitem{noack2025ai}
Noack, C.A., Mackensen, J., Guhl, J., Zowalla, R., Bienzeisler, B., Neuh{\"u}ttler, J.: [ai] deation: Genai-based collaborative service innovation. Proceedings of the 58th Hawaii International Conference on System Sciences pp. 1336--1345 (2025)

\bibitem{moeka_towards_2024}
Nomura, M., Ito, T., Ding, S.: Towards collaborative brain-storming among humans and ai agents: An implementation of the ibis-based brainstorming support system with multiple ai agents. In: Proceedings of the ACM Collective Intelligence Conference. p. 1–9. CI '24, Association for Computing Machinery, New York, NY, USA (2024). \doi{10.1145/3643562.3672609}

\bibitem{shin2022chatbots}
Shin, J., Hedderich, M.A., Lucero, A., Oulasvirta, A.: Chatbots facilitating consensus-building in asynchronous co-design. In: Proceedings of the 35th Annual ACM Symposium on User Interface Software and Technology. pp. 1--13. Association for Computing Machinery, New York, NY, USA (2022). \doi{10.1145/3526113.3545671}

\bibitem{Shortall_2022}
Shortall, R., Itten, A., Meer, M.v.d., Murukannaiah, P., Jonker, C.: Reason against the machine? future directions for mass online deliberation. Frontiers in Political Science  \textbf{4},  946589 (Oct 2022). \doi{10.3389/fpos.2022.946589}

\bibitem{Tessler_common_ground}
Tessler, M.H., Bakker, M.A., Jarrett, D., Sheahan, H., Chadwick, M.J., Koster, R., Evans, G., Campbell-Gillingham, L., Collins, T., Parkes, D.C., Botvinick, M., Summerfield, C.: Ai can help humans find common ground in democratic deliberation. Science  \textbf{386}(6719),  eadq2852 (2024). \doi{10.1126/science.adq2852}

\bibitem{wan_second_mind}
Wan, Q., Hu, S., Zhang, Y., Wang, P., Wen, B., Lu, Z.: "it felt like having a second mind": Investigating human-ai co-creativity in prewriting with large language models. Proc. ACM Hum.-Comput. Interact.  \textbf{8}(CSCW1) (Apr 2024). \doi{10.1145/3637361}

\bibitem{zarate2013collaborative}
Zarat{\'e}, P., Konat{\'e}, J., Camilleri, G.: Collaborative decision making tools: A comparative study based on functionalities. In: Proceedings of the 13th International Conference Group Decision and Negotiation Part III. pp. 111--122 (2013)

\bibitem{zha2024designing}
Zha, S., Qiao, Y., Hu, Q., Li, Z., Gong, J., Xu, Y.: Designing child-centric ai learning environments: Insights from llm-enhanced creative project-based learning (2024), \url{https://arxiv.org/abs/2403.16159}

\end{thebibliography}
\end{document}